\documentclass[pra,twocolumn,showpacs,preprintnumbers,amsmath,amssymb]{revtex4}
\usepackage{graphicx}% Include figure files
\usepackage{dcolumn}% Align table columns on decimal point
\usepackage{bm}% bold math

\newcommand{\vect}[1]{{\bf #1}}
\newcommand{\aho}{a_\mathrm{HO}}

\newcommand{\rr}{{\bf r}}
\newcommand{\ti}[1]{\tilde{#1}}
\newcommand{\LL}{\mathcal{L}}
\newcommand{\HH}{\hat{\mathcal{H}}}
\newcommand{\integ}[1]{\langle #1 \rangle}
\begin{document}

\preprint{APS/123-QED}

\title{Splitting Instability of a Multiply Charged Vortex in a Bose-Einstein Condensate}
\author{Yuki Kawaguchi}
 \email{yuki@scphys.kyoto-u.ac.jp}
\author{Tetsuo Ohmi}
\affiliation{Department of Physics, Graduate School of Science, Kyoto University, Kyoto 606-8502, Japan}
\date{\today}% It is always \today, today,

\begin{abstract}

We consider the splitting mechanism of a multiply charged vortex into singly charged vortices
in a Bose-Einstein condensate confined in a harmonic potential at zero temperature.
The Bogoliubov equations support unstable modes with complex eigenfrequencies (CE modes),
which cause the splitting instability without the influence of thermal atoms.
The investigation of the excitation spectra shows that
the negative-energy (NE) mode plays an important role in the appearance of the CE modes.
The configuration of vortices in splitting is determined by the angular momentum of the associated NE mode.
This structure has also been confirmed by the numerical simulation of the time-dependent Gross-Pitaevskii equation.

\end{abstract}

%\keywords{ keywords}%Use showkeys class option if keyword
\maketitle

\section{INTRODUCTION}
The discovery of Bose-Einstein condensates (BECs) of alkali-metal atomic gases
enables us to make fundamental investigations into superfluidity and quantized vortices.
Vortices in BECs have been created in three ways:
phase imprinting~\cite{Matthews1999,Williams1999}, optical spoon stirring~\cite{Madison2000}
and topological phase engineering~\cite{Isoshima2000,Leanhardt2002}.
By using the third method, multiply charged singular vortices are created in BECs,
which have not been achieved in other systems such as superfluid $^3$He, $^4$He~\cite{foot1,foot2}.

Recently, MIT group succeeded in imprinting vortices in Bose-Einstein condensates by using topological phases~\cite{Leanhardt2002}.
They observed doubly and quadruply quantized vortices in a non-rotating trap.
Although such a multiply charged vortex is expected to be unstable and split into singly charged vortices,
no obvious splitting has been observed in the experiment.
Unfortunately, since the center of the external potential changes from a minimum to a saddle in the vortex imprinting process,
the vortex state cannot be trapped.
Therefore, the fate of the vortices has not been revealed in the experiment.

In a non-rotating trap, the energy of a vortex state increases in proportion to the square of the winding number~\cite{nozieres-pines}.
When there are some dissipative processes such as scattering with thermal atoms,
the multiply charged vortex should split into singly quantized vortices and escape from the condensate.
%The time scale depends on the dissipation.
The existence of the negative energy eigenvalues among the excitation spectra shows that
the vortex state continuously turns into a vortex-free state without any energy barriers~\cite{Rokhsar1997,Svidzinsky1998,Svidzinsky2000}.
Then, what happens in the system in which all atoms are condensed and scattering with thermal atoms can be negligible?
Does the vortex remain in the condensate without splitting?

There is another candidate for splitting instability.
For a multiply charged vortex in a BEC,
the Bogoliubov equations possess complex eigenfrequencies
in certain regions of the parameter space of the interaction strength~\cite{Pu1999,Skryabin2000,Mottonen2003}.
This fact means that there is an eigenmode growing exponentially and
the condensate becomes unstable against infinitesimal perturbations.
Therefore the vortex state is intrinsically unstable.

In this paper,
we consider the origin of the complex-eigenfrequency (CE) modes and their effect upon the vortex splitting process.
A vortex with a winding number $L>2$ may split into various states,
while a doubly charged vortex merely splits into a pair of singly charged vortices~\cite{Mottonen2003}.
We will treat a quadruply quantized vortex in concrete calculations in this paper.
In Sec.~\ref{sec:formalism}, we briefly introduce the formalisms based on the mean field theory.
The orthonormal conditions for CE modes are constructed there.
In Sec.~\ref{sec:spectrum}, a clear insight into the origin of CE and negative-energy (NE) modes is provided by
the collective excitation spectra of the Bogoliubov equation.
The possible structures of splitting are also considered in Sec.~\ref{sec:sim},
showing the results of numerical simulations of the Gross-Pitaevskii equation (GPE).

\section{\label{sec:formalism}FORMALISM}
\subsection{Stationary state with a multiply charged vortex}
BECs of weakly interacting atoms are well described with the GPE given by
\begin{eqnarray}
 i\hbar\frac{\partial \Psi}{\partial t} = \left(-\frac{\hbar^2}{2M}\nabla^2 + V_{\rm tr}(\rr) - \mu + g|\psi|^2\right)\Psi,
\end{eqnarray}
where $\Psi(\rr,t)$ is a condensate wave function, $M$ is the mass of the trapped atoms,
$\mu$ is the chemical potential, and $g=4\pi\hbar^2a/M$ represents the strength of the interparticle interaction.
We treat repulsively interacting atoms and assume the $s$-wave scattering length $a$ to be positive.
An external trap potential has an axially symmetrical form with trap frequencies $\omega_r$ and $\omega_z$ as
$V_{\rm tr}(\rr)=\frac{1}{2}M\omega_r^2(x^2+y^2)+\frac{1}{2}M\omega_z^2z^2$.
For simplicity, we assume a pancake-shaped BEC, i.e. $\omega_r\ll \omega_z$,
and fix the $z$-dependence of the orderparameter as $\Psi(\rr,t)\equiv\psi(x,y,t)\psi_z(z)$,
where $\psi_z(z)$ is the normalized ground state wave function of a one-dimensional harmonic oscillator.
Then we treat the two-dimensional wave function $\psi(x,y,t)$ which is normalized to the total atom number $N$.

%BECs of weakly interacting atoms are well described with the GPE.
%We assume a pancake-shaped BEC for simplicity, and consider a two-dimensional system.
%A condensate wave function $\psi(x,y,t)$ obeys the 2D-GPE given by
%\begin{eqnarray}
% i\hbar\frac{\partial \psi}{\partial t} = \left(-\frac{\hbar^2}{2M}\nabla^2 + V_{\rm tr}(\rr) - \mu + g|\psi|^2\right)\psi,
%\end{eqnarray}
%where $M$ is the mass of the trapped atoms, $\nabla^2=\partial_x^2+\partial_y^2$, $\mu$ is the chemical potential,
%and $g=4\pi\hbar^2a/M$ represents the strength of the interparticle interaction.
%We treat repulsively interacting atoms and assume the $s$-wave scattering length $a$ to be positive.
%An external trap potential has an axially symmetrical form with a radial trap frequency $\omega_r$ as
%$V_{\rm tr}(\rr)=\frac{1}{2}M\omega_r^2(x^2+y^2)$.
%The wave function is normalized so that the number of atoms $N_z$ per unit length along $z$-axis is
%$\int\!\!\!\int dxdy|\psi|^2=N_z$.

It is convenient to introduce the dimensionless time and space variables $\ti{t}=\omega_rt$ and $(\ti{x},\ti{y})=(x,y)/\aho$,
and the normalized order parameter $\ti{\psi}=\aho\psi/\sqrt{N}$,
where $\aho$ is the harmonic oscillator length: $\aho=\sqrt{\hbar/2M\omega_r}$.
Energies are also scaled by $\hbar \omega_r$, for example, $\ti{\mu}=\mu/\hbar\omega_r$.
Then the GPE is rewritten in a simple form
\begin{eqnarray}
i\frac{\partial \psi}{\partial t}=\left(-\nabla^2+\frac{1}{4}r^2-\mu+\eta|\psi|^2\right)\psi,
\label{eq:GP}
\end{eqnarray}
where $\nabla^2=\partial_x^2+\partial_y^2, r^2=x^2+y^2, \eta=8\pi a N$ and we omit tilde for simplicity.

Assuming an axial symmetry around a vortex line,
we express the condensate wave function with an $L$-charged vortex in equilibrium as
\begin{eqnarray}
\psi_0(r,\theta)=A(r)e^{iL\theta},\hspace{2mm}L=0,1,2,\cdots,
\label{eq:equil}
\end{eqnarray}
where $A(r)$ is a real function and satisfies the equations
\begin{align}
 &\left[\LL(L)+\eta A(r)^2\right]A(r)=0, \label{eq:GP-A}\\
 &\LL(L)\equiv-\frac{d^2}{dr^2}-\frac{1}{r}\frac{d}{dr}+\frac{L^2}{r^2}+\frac{1}{4}r^2-\mu.
\end{align}
The chemical potential $\mu$ is found from the normalization condition
$2\pi\int_0^\infty rdr\left\{A(r)\right\}^2=1$.
We assume $L$ to be positive without loss of generality.
The results of numerical calculations for $L=4$ are presented in Secs.~\ref{sec:spectrum} and \ref{sec:sim}.

\subsection{Collective excitations}
To study the collective excitations of a BEC,
we add small fluctuations to the stationary state as
$\psi(r,\theta,t)=\left[A(r)+f(r,\theta,t)\right]e^{iL\theta}$,
where $f$ is small and complex.
Assuming that the excitations are periodic in $\theta$ with period $2\pi$,
we expand $f$ into a Fourier series:
\begin{eqnarray}
f(r,\theta,t)=\sum_{l=0,1,2,\cdots}\left[u_l(r,t)e^{il\theta}+v_l^\ast(r,t)e^{-il\theta}\right].
\label{eq:fluc}
\end{eqnarray}
%Here $\pm l$ correspond to the angular momenta relative to the condensate.
Substituting $\psi(r,\theta,t)$ to Eq.~\eqref{eq:GP} and linearizing with respect to $f$,
one obtains the well-known Bogoliubov equation
\begin{eqnarray}
\HH_l \vect{w}_l(r)=\omega_l \hat{\sigma}\vect{w}_l(r),
\label{eq:BE}
\end{eqnarray}
where
\begin{eqnarray}
\vect{w}_l(r)=\left(\begin{array}{c c}u_l(r) \\ v_l(r) \end{array}\right), \hspace{5mm}
 \hat{\sigma}=\left(\begin{array}{c c} 1 & 0 \\ 0 & -1 \end{array}\right),
\end{eqnarray}
and $\HH_l$ is a symmetrical matrix
\begin{eqnarray}
\HH_l=
\left(\begin{array}{c c}
\LL(L+l)+2\eta A^2 & \eta A^2 \\
\eta A^2 & \LL(L-l)+2\eta A^2
\end{array}\right).
\end{eqnarray}
We are interested in the eigenstates of $\hat{\sigma}\HH_l$ and take the time dependence of the excitations as
$\vect{w}_l(r,t)=\vect{w}_l(r)\exp(-i\omega_l t)$.
It should be noted here that the Bogoliubov equations with different $l$ are independent.

%The fluctuations with angular momenta $\pm l$ are coupled to each other,
%while the Bogoliubov equations with different $|l|$ are independent.
Usually, the summation in Eq.~\eqref{eq:fluc} is done for both positive and negative $l$
which correspond to the angular momenta of excitations,
and the normalization condition $\langle|u_l|^2-|v_l|^2\rangle=1(>0)$ is imposed,
where $\integ{\cdots}\equiv 2\pi\int_0^\infty \cdots rdr$.
In this case, however, we remove this normalization condition and restrict $l\ge0$,
since the eigenvalue equation for $-l$ is quite the same as those of $l$
when one replaces $u_l,v_l,\omega_l$ with $v_{-l},u_{-l}$, and $-\omega_{-l}$, respectively.
Then we define the angular momentum $l_{\rm ex}$ of excitations by comparing the amplitudes of components $\exp(\pm il\theta)$ as
\begin{eqnarray}
l_{\rm ex}=l\times{\rm sgn}\integ{|u_l|^2-|v_l|^2}.
\label{eq:am}
\end{eqnarray}
%Although $l<0$ are often used for describing negative-angular-momentum excitations,
%they yield quite the same eigenvalue equations as those of $-l\ (>0)$
%when one replaces $u_l,v_l,\omega_l$ with $v_{-l},u_{-l}$, and $-\omega_{-l}$, respectively.
%Here we restrict $l\ge 0$ and treat the excitations with both positive and negative angular momenta.
To specify the sign of the angular momenta,
we use $\vect{w}_l^{\sf (u)}$ and $\vect{w}_l^{\sf (v)}$ for the eigenfunctions of $l_{\rm ex}=l$ and $l_{\rm ex}=-l$, respectively.
The excitation energy is also defined by
\begin{eqnarray}
 \epsilon_l=\omega_l\times\integ{|u_l|^2-|v_l|^2},
  \label{eq:ene}
\end{eqnarray}
reflecting the relation $\omega_l=-\omega_{-l}$.
This is confirmed by expanding the Hamiltonian up to the second order in perturbations.

\subsection{Complex eigenmodes and orthonormality}
Equation~\eqref{eq:BE} is allowed to have complex eigenfrequencies since it is a non-Hermite eigenvalue equation~\cite{Pu1999}.
Here, we derive the normalization and orthogonality conditions for eigenmodes including CE modes.

Consider two eigenmodes $\vect{w}_{lm,n}$ of Eq.~\eqref{eq:BE} with eigenfrequencies $\omega_{lm,n}$.
Using the property of the Hermitian operator $\HH_l$:
$\integ{\vect{w}_{ln}^T\HH_l\vect{w}_{lm}}=\integ{\vect{w}_{lm}^T\HH_l\vect{w}_{ln}}$,
where $^T$ denotes the transpose, one obtains the equation
\begin{eqnarray}
(\omega_{lm}-\omega_{ln})\integ{\vect{w}_{ln}^T\hat{\sigma}\vect{w}_{lm}}=0.
\label{eq:orthonormal}
\end{eqnarray}

When two eigenstates have different eigenfrequencies ($\omega_{lm}\neq\omega_{ln}$),
they are orthogonal: $\integ{\vect{w}_{ln}^T\hat{\sigma}\vect{w}_{lm}}=0$.
Especially, in case when $\vect{w}_{l\mu}$ is complex with a complex eigenvalue $\omega_{l\mu}$,
its complex conjugate $\vect{w}_{l\mu}^\ast$ is also an eigenstate with eigenvalue $\omega_{l\mu}^\ast$.
The orthogonality condition for $\vect{w}_{l\mu}$ and $\vect{w}_{l\mu}^\ast$ is written as
\begin{eqnarray}
\integ{\vect{w}^\dagger_{l\mu}\hat{\sigma}\vect{w}_{l\mu}}=\integ{|u_{l\mu}|^2-|v_{l\mu}|^2}=0.
\label{eq:complex_orth}
\end{eqnarray}
According to Eqs.~\eqref{eq:am} and \eqref{eq:ene}, a CE mode corresponds to the excitation with
zero angular momentum and zero excitation energy.

Equation~\eqref{eq:orthonormal} indicates that $\integ{\vect{w}_{ln}^T\hat{\sigma}\vect{w}_{ln}}$ can be normalized.
For a real-eigenfrequency (RE) mode, we choose the phase so that the eigenfunction is real.
Then the normalization constant is defined to be consistent with its angular momentum as
\begin{eqnarray}
\integ{\vect{w}_{ln}^{{\sf (u)} T}\hat{\sigma}\vect{w}_{ln}^{\sf (u)}}=1,\hspace{2mm}
\integ{\vect{w}_{ln}^{{\sf (v)} T}\hat{\sigma}\vect{w}_{ln}^{\sf (v)}}=-1.
\end{eqnarray}
For a CE mode, 
we divide the eigenfunction into two real functions $\vect{w}^{\sf (R,I)}$
as $\vect{w}_{l\mu}=\frac{1}{\sqrt{2}}(\vect{w}^{\sf (R)}+ i\vect{w}^{\sf (I)})$.
The phase of $\vect{w}_{l\mu}$ is determined in such a way that the following conditions are satisfied,
\begin{subequations}
\begin{align}
&\integ{\vect{w}^{{\sf (R)}T}\hat{\sigma}\vect{w}^{\sf (R)}}=1,\hspace{2mm}
\integ{\vect{w}^{{\sf (I)}T}\hat{\sigma}\vect{w}^{\sf (I)}}=-1, \label{eq:complex_norm} \\
&\integ{\vect{w}^{{\sf (R)}T}\hat{\sigma}\vect{w}^{\sf (I)}}=0.
\end{align}
\end{subequations}
With these conditions, the CE mode indeed satisfies Eq.~\eqref{eq:complex_orth} and the normalization conditions
$\integ{\vect{w}_{l\mu}^T\hat{\sigma}\vect{w}_{l\mu}}
=\integ{(\vect{w}_{l\mu}^\ast)^T\hat{\sigma}\vect{w}_{l\mu}^\ast}=1$.
According to Eq.~\eqref{eq:complex_norm}, the function $\vect{w}^{\sf (R,I)}$ correspond to
positive- and negative-angular-momentum excitations (not eigenstates).
We use the expression $\vect{w}_{lm'}^{\sf (u)}$ and $\vect{w}_{ln'}^{\sf (v)}$ instead of $\vect{w}^{\sf (R,I)}$, respectively,
and treat them and other eigenmodes equally.

In this way, the orthonormal basis set $\vect{w}_{ln}^{\sf (u,v)}$ has been constructed with the following conditions:
\begin{subequations}
\begin{align}
&\integ{\vect{w}_{lm}^{{\sf (u)}T}\hat{\sigma}\vect{w}_{ln}^{\sf (u)}}=\delta_{mn},\hspace{2mm}
\integ{\vect{w}_{lm}^{{\sf (v)}T}\hat{\sigma}\vect{w}_{ln}^{\sf (v)}}=-\delta_{mn},\\
&\integ{\vect{w}_{lm}^{{\sf (u)}T}\hat{\sigma}\vect{w}_{ln}^{\sf (v)}}=0.
\end{align}
\label{eq:all_orthonorm}
\end{subequations}

\section{\label{sec:spectrum}ENERGY SPECTRUM}
\subsection{Collective excitations with $\eta=0$}
First, we investigate at non-interacting limit ($\eta=0$).
The equilibrium state satisfies  $\LL(L)A(r)=0$.
The eigen function $F_n^L$ and eigenvalue $E_n^L$  are given analytically by
\begin{eqnarray}
F_n^L(r)&=& C_n^L \  r^{|L|}\ \mathfrak{L}_n^{|L|}(r^2/2) \  e^{-r^2/4},\\
E_n^L&=&|L|+2n+1,
\end{eqnarray}
where $n=0, 1, 2, \cdots$ denotes the radial quantum number and
 $\mathfrak{L}_n^L(x)$ is the generalized Laguerre polynomial function.
The coefficient $C_n^L$ is defined by the normalization condition $\integ{F_n^L(r)^2} = 1$.
The stationary state $A_0(r)$ and its chemical potential $\mu_0$ are
\begin{eqnarray}
A_0(r)=F_0^L(r),\hspace{2mm}\mu_0=E_0^L=L+1.
\end{eqnarray}

\begin{figure}
\includegraphics[width=8cm]{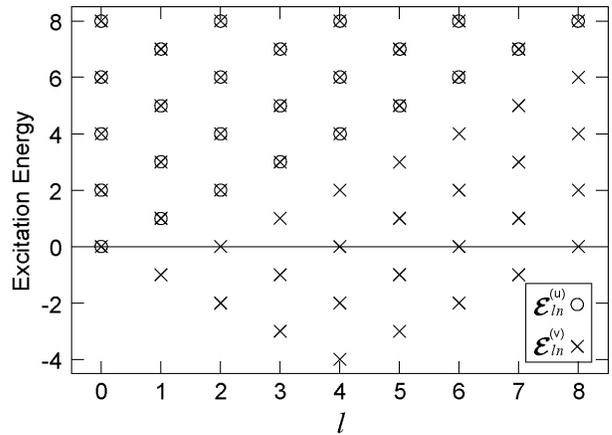}
\caption{\label{fig:spectra_noint}Excitation spectra of a quadruply charged vortex state at non-interacting limit.
The symbols $\circ$ and $\times$ represent $\mathcal{E}_{ln}^{\sf (u)}$ and $\mathcal{E}_{ln}^{\sf (v)}$, respectively.}
\end{figure}

The Bogoliubov equation with $\eta=0$ decouples into the independent equations
\begin{subequations}
\begin{align}
 \LL(L+l)U_l&=\Omega_l U_l,\\
 \LL(L-l)V_l&=-\Omega_l V_l,
\end{align}
\end{subequations}
where we write eigenfunctions and eigenfrequencies with $\eta=0$ in capital letters.
The solutions are $U_l=F_n^{L+l}, V_l=F_n^{L-l}$, 
i.e. the set of eigenmodes is given by
\begin{eqnarray}
\vect{W}_{ln}^{\sf (u)}=F_n^{L+l}(r)
\left(\begin{array}{c} 1 \\ 0 \end{array}\right),\hspace{2mm}
\vect{W}_{ln}^{\sf (v)}=F_n^{L-l}(r)
\left(\begin{array}{c} 0 \\ 1 \end{array}\right),
\end{eqnarray}
having eigenfrequencies $\Omega_{ln}^{\sf (u)}=|L+l|+2n-L$ and $\Omega_{ln}^{\sf (v)}=-|L-l|-2n+L$, respectively.
According to Eq.~\eqref{eq:ene},
the excitation energy of $\vect{W}_{ln}^{\sf (v)}$ is $\mathcal{E}_{ln}^{\sf (v)}=-\Omega_{ln}^{\sf (v)}$,
while that of $\vect{W}_{ln}^{\sf (u)}$ is $\mathcal{E}_{ln}^{\sf (u)}=\Omega_{ln}^{\sf (u)}$.
Figure~\ref{fig:spectra_noint} shows the excitation spectra with $\eta=0$ with respect to the equilibrium state
with a quadruply quantized vortex ($L=4$).
There are a few negative energies among the spectra of $1\le l\le 2L-1$.
These NE modes come from the fact that $\psi_0$ is not a ground state.
For a non-interacting BEC, NE modes correspond to the eigenstates in the trap with lower energies than $\psi_0$.
The existence of these various NE modes shows how unfavorable the multiply charged vortex is.

%The vortex state with finite $\eta$ has higher energy than vortex-free state, and therefore NE modes always exist.
%In the case of a dissipative system, the NE modes are initially excited in the process in which the vortex state turns into the vortex-free state.
In the case of a non-isolated system, the energy of the condensate does not need to be conserved.
It decreases through the growth of NE modes, leading to the vortex decay.
In an isolated system, however, the energy must be conserved
and the excitation which can cause the vortex decay is not a NE mode but a CE mode,
as we will show in the following.

\subsection{$\eta$-dependence of energy levels}
To calculate the energy spectrum for $\eta > 0$,
we expand the eigenmodes $\vect{w}_l$ in $\vect{W}_{ln}^{\sf (u,v)}$ as
\begin{eqnarray}
\vect{w}_l
=\sum_{n=0}^N(\alpha_n \vect{W}_{ln}^{\sf (u)}+\beta_n\vect{W}_{ln}^{\sf (v)}).
\label{eq:w_expand}
\end{eqnarray}
Substituting Eq.~\eqref{eq:w_expand} to Eq.~\eqref{eq:BE}, we rewrite the Bogoliubov equation as eigenvalue equations for $(\alpha_0,\beta_0,\cdots,\alpha_N,\beta_N)$:
\begin{subequations}
\begin{eqnarray}
\eta\sum_{m=0}^N\left[2\integ{F_n^{L+l}A^2F_m^{L+l}}\alpha_m+\integ{F_n^{L+l}A^2F_m^{L-l}}\beta_m \right] \nonumber \\
+(E_n^{L+l}-\mu-\omega_l)\alpha_n=0,\ \ \ \\
\eta\sum_{m=0}^N\left[\integ{F_n^{L-l}A^2F_m^{L+l}}\alpha_m+2\integ{F_n^{L-l}A^2F_m^{L-l}}\beta_m\right]\nonumber\\
+(E_n^{L-l}-\mu+\omega_l)\beta_n=0,\ \ \ 
\end{eqnarray}
\label{eq:BE2}\end{subequations}
where $n=0, 1, \cdots, N$ and $A(r), \mu$ are the numerical solutions of Eq.~\eqref{eq:GP-A}.

\begin{figure}[tbp]
\includegraphics[width=8cm]{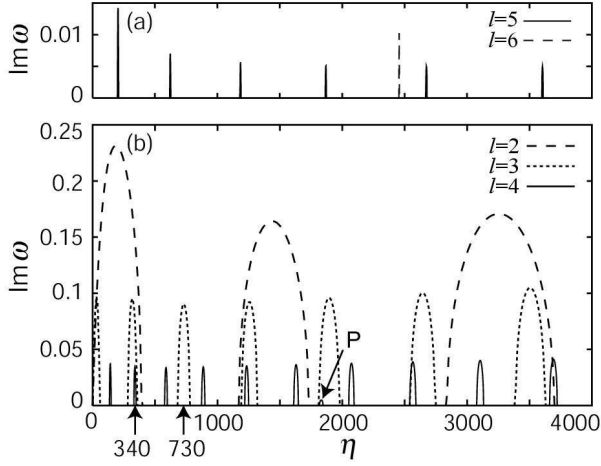}
\caption{\label{fig:im}Imaginary parts of eigenfrequencies for (a) $l=5,6$ and (b) $l=2,3,4$ as functions of the interaction strength $\eta$.
The result for $l=6$ is magnified a hundred times.
The CE mode at {\sf P} differs from others with $l=4$ as to the associated NE mode (see text).}
\end{figure}

We have numerically solved Eq.~\eqref{eq:BE2} for $L=4$ and $0\le l\le 8$ in the range of $0 \le \eta \le 4000$.
The number of terms summed in Eq.~\eqref{eq:BE2} is $N=50$.
The NE modes exist for $1 \le l \le 7$ as expected from the result of $\eta=0$.
The CE modes appear for $2\le l \le6$.
The imaginary parts of complex eigenfrequencies are shown in Fig.~\ref{fig:im},
which agrees well with the results of doubly and triply charged vortices~\cite{Pu1999}.
\begin{figure}[tbp]
\includegraphics[width=8cm]{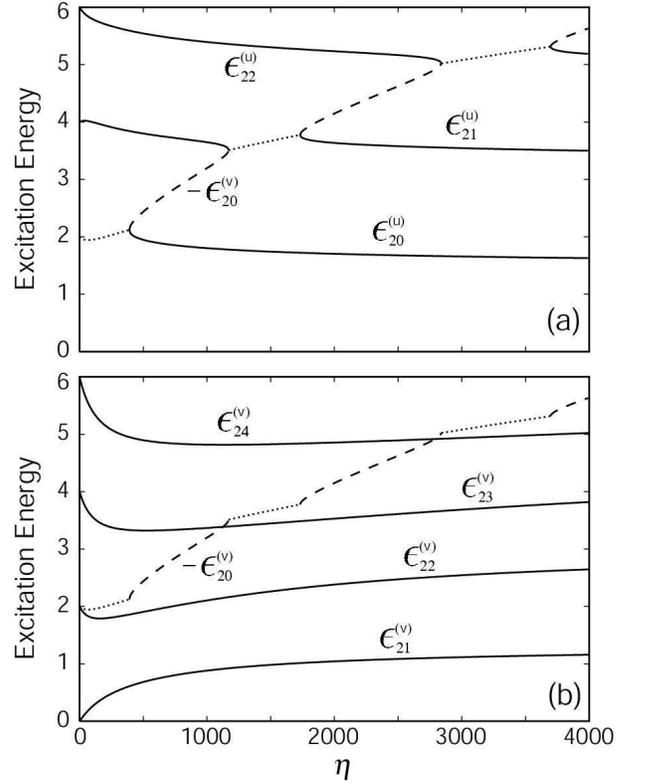}
\caption{\label{fig:l2}Excitation spectra for $l=2$.
The solid lines correspond to excitation energies with (a) positive angular momenta and (b) negative angular momenta.
In both figures, the negative energy eigenvalues and the real part of the complex eigenfrequency are plotted by broken and dotted lines, respectively}
\end{figure}

The origin of the CE mode is made clear
by plotting them as functions of $\eta$.
Figure~\ref{fig:l2} shows the excitation spectra with $l=2$.
Each eigenstate is identified by a radial quantum number $n=0,1,2,\cdots$ and a sign of angular momentum ${\sf u,v}$.
The solid lines in Figs.~\ref{fig:l2}(a) and (b) represent $\epsilon_{2n}^{\sf (u)}$ and $\epsilon_{2n}^{\sf (v)}$, respectively.
There is a NE mode $\vect{w}_{20}^{\sf (v)}$, and $-\epsilon_{20}^{\sf (v)}(>0)$ is plotted with a broken line in both figures.
We also plot the real parts of the complex frequencies with dotted lines.
It is clear from Fig.~\ref{fig:l2} that
a conjugate pair of CE modes appears instead of two RE modes $\vect{w}_{2n}^{\sf (u)}$ and $\vect{w}_{20}^{\sf (v)}$
in the regions where $\epsilon_{2n}^{\sf (u)}+\epsilon_{20}^{\sf (v)}=0$.
As we mentioned in Sec.~\ref{sec:formalism}, a CE mode corresponds to an excitation with zero excitation energy and zero angular momentum,
which are made up of two excitations such as $\frac{1}{\sqrt{2}}(\vect{w}_{2n}^{\sf (u)}\pm i \vect{w}_{20}^{\sf (v)})$.
The real and imaginary parts, which are divided so as to satisfy the orthonormality, have a physical meaning:
they correspond to positive- and negative-angular-momentum excitations,
which continuously change to RE modes as $\eta$ changes.

The second quantized description simplifies this picture~\cite{Fetter1972,Pu1999}.
A boson field operator is defined as $\hat{\psi}(\rr)=\psi_0(\rr)+\hat{\phi}(\rr)$,
where the {\it c}-number function $\psi_0(\rr)$ denotes the condensate wave function and $\hat{\phi}(\rr)$ is the fluctuation part.
We decompose $\hat{\phi}(\rr)$ as
$\hat{\phi}(\rr)=e^{iL\theta}\sum_{\lambda,l,n}
[u_{ln}^{(\lambda)}(r)e^{il\theta}\alpha_{\lambda ln}+v_{ln}^{(\lambda)}(r)e^{-il\theta}\alpha_{\lambda ln}^\dagger]$.
Here, $\alpha_{\lambda ln}^\dagger$ is a creation operator associated with a fluctuation $\vect{w}_{ln}^{\sf (\lambda)}$.
Let us fix the interaction strength, for example, to $\eta=200$ where a pair of the CE modes
$\frac{1}{\sqrt{2}}(\vect{w}_{20}^{\sf (u)}\pm i\vect{w}_{20}^{\sf (v)})$ exists.
Then the Bogoliubov Hamiltonian is written as
\begin{align}
 \hat{\mathcal{K}}_B=
 &\sum_{\lambda={\sf u,v}}\sum_{(l,n)\neq(2,0)} \epsilon_{ln}^\lambda\alpha_{\lambda ln}^\dagger\alpha_{\lambda ln} \nonumber \\
 &+ {\rm Re}\,\omega_c \left(\alpha_{{\sf u}20}^\dagger\alpha_{{\sf u}20}-\alpha_{{\sf v}20}^\dagger\alpha_{{\sf v}20}\right) \nonumber\\
 &+ {\rm Im}\,\omega_c \left(\alpha_{{\sf u}20}^\dagger\alpha_{{\sf v}20}^\dagger+\alpha_{{\sf u}20}\alpha_{{\sf v}20}\right),
\label{eq:hamil}
\end{align}
where $\omega_c$ is the complex eigenfrequency, and we have used the relations
\begin{subequations}
\begin{align}
{\rm Re}\,\omega_{c}&=\integ{\vect{w}_{20}^{{\sf(u)} T}\HH_2\vect{w}_{20}^{\sf (u)}}
=-\integ{\vect{w}_{20}^{{\sf(v)} T}\HH_2\vect{w}_{20}^{\sf (v)}},\\
{\rm Im}\,\omega_{c}&=\integ{\vect{w}_{20}^{{\sf(u)} T}\HH_2\vect{w}_{20}^{\sf (v)}}.
\end{align}
\end{subequations}
Equation~\eqref{eq:hamil} clearly shows that the two excitations $\vect{w}_{20}^{\sf (u,v)}$ are coupled to each other
and should be created or annihilated together.

\begin{figure}
\includegraphics[width=8cm]{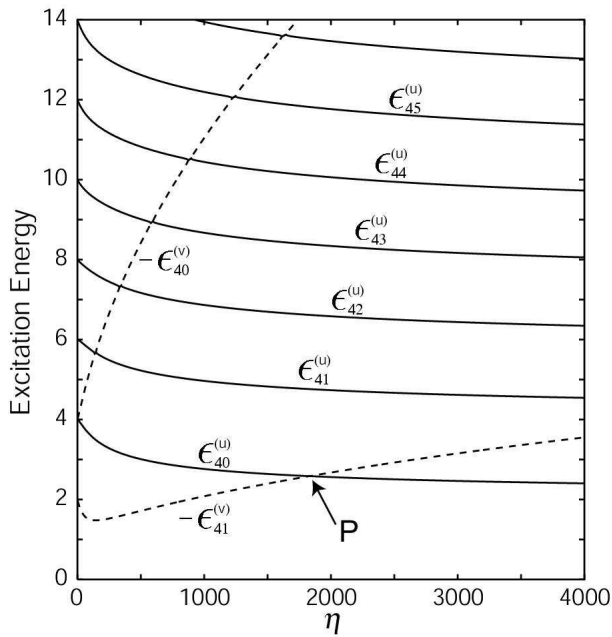}
\caption{\label{fig:l4} Excitation spectra for $l=4$.
The description of lines are in Fig.~\ref{fig:l2}(a).
There are two NE modes and both turn into CE excitations.
The point {\sf P} corresponds to {\sf P} in Fig.~\ref{fig:im} (see text).}
\end{figure}

In the above discussion, the NE mode is important for the appearance of CE modes.
In the case of $l=2$, there exist only one NE mode, $\vect{w}_{20}^{\sf (v)}$.
Therefore, all CE modes with $l=2$ are found by tracing $\epsilon_{20}^{\sf (v)}$.
We also present the spectra for $l=4$ in Fig.~\ref{fig:l4}, where the solid lines represent $\epsilon_{4n}^{\sf (u)}$.
There exist two negative energy eigenvalues $\epsilon_{40}^{\sf (v)}$ and $\epsilon_{41}^{\sf (v)}$ which are plotted by broken lines.
We have confirmed that the NE mode $\vect{w}_{41}^{\sf (v)}$, as well as $\vect{w}_{40}^{\sf (v)}$, turns into
CE modes in the region where $\epsilon_{40}^{\sf (u)}+\epsilon_{41}^{\sf (v)}=0$ is satisfied
(at point {\sf P} in Figs.~\ref{fig:im}(b) and \ref{fig:l4}).
We have obtained all of the CE modes by inspecting the behaviors of NE modes (see Fig.~\ref{fig:im}).

The $\eta$-dependence of negative energy eigenvalues is also important in this mechanism.
Since their dependence differs from that of positive energies,
many CE modes appear (Figs.~\ref{fig:l2}(a) and \ref{fig:l4}).
The qualitative account for the $\eta$-dependence can be given by introducing the effective potential
\begin{eqnarray}
 V_{\rm eff}(l,r)=\frac{1}{4}r^2+\frac{(L-l)^2}{r^2}+2\eta A(r)^2-\mu,
\end{eqnarray}
which is easily derived from the Bogoliubov equation for negative-angular-momentum ($-l$) excitations.
When the equilibrium state contains a vortex, the $A^2$ term produces a potential well centered at the vortex core~\cite{Svidzinsky1998}.
On the other hand, the second term represents the centrifugal potential depending upon the angular momentum of the fluctuation.
This term makes the well narrower as $|L-l|$ increases.
A NE mode corresponds to a bound state of the well, and therefore shows different $\eta$-dependence from others.

There are two effects on the negative energy eigenvalues associated with the change of $\eta$.
By increasing $\eta$, (i) the chemical potential $\mu$ increases as well, leading to a deeper well and smaller energy eigenvalues and
(ii) the coherence length (i.e. vortex core size) decreases, leading to a narrower well and larger energy eigenvalues.
In the case of $l=4(=L)$, the negative energy eigenvalues decrease rapidly as $\eta$ increases by (i).
In contrast, $l=1$ negative energy eigenvalue increases due to (ii).
Therefore, the condition $\epsilon_{1n}^{\sf (u)}+\epsilon_{10}^{\sf (v)}=0$ cannot be satisfied with any eigenmodes $\vect{w}_{1n}^{\sf (u)}$,
and no CE mode appears.

\section{\label{sec:sim}PATTERNS OF VORTEX SPLITTING}
So far our discussion has been based on the spectrum analysis.
We have found that the CE mode corresponds to a zero-energy excitation and should grow in an isolated system.
In this section we consider the possible patterns of vortex splitting.

\subsection{Vortex structure with elementary excitations}
Here, we analyze the structure of BEC with fluctuations.
The wave function perturbed by an eigenmode $\vect{w}_l$ is written as
$\psi=\left[A(r)+\delta\left(u_le^{il\theta}+v_l^\ast e^{-il\theta}\right)\right]e^{iL\theta}$,
where $\delta$ is a small and real constant.

The structure of the vortex core in equilibrium can be described with an asymptotic form
$A(r)\sim r^L\ (r\to 0)$.
The eigenmode at the center also behaves as
$u_l\sim r^{|L+l|}$ and $v_l\sim r^{|L-l|}$.
Then the perturbed wave function is written in the lowest order of $r$ as
$\psi\sim \delta\ r^{|L-l|}e^{i(L-l)\theta}$,
i.e. an $(L-l)$-charged vortex exists at the center.
One also finds that the condensate has $l$-fold symmetry
by linearizing the density with respect to $\delta$:
\begin{eqnarray}
|\psi(r,\theta)|^2=A^2+ 2A\delta \left[{\rm Re}(u_l+v_l)\cos(l\theta)\right.\ \ \ \ \ \nonumber\\
\left.-{\rm Im}(u_l+v_l)\sin(l\theta)\right].
\end{eqnarray}
Assuming the conservation of the total angular momentum,
splitting of an $L$-charged vortex obeys the rules:
(i) an $(L-l)$-charged vortex stays at the trap center, and (ii) $l$ singly quantized vortices are arranged in $l$-fold symmetry. 

Although the above rules are applicable to both RE and CE modes,
only the latter can grow exponentially and cause large changes of the condensate with a small perturbation.
In the following, we investigate the effect of CE modes to the time development of a vortex state.

\subsection{Projection to elementary excitations}
We solve the GPE numerically and simulate the time development of splitting process.
Regarding the time evolution of a condensate as individual developments of elementary excitations,
we again expand the time-depending wave function as
\begin{eqnarray}
 \psi(r,\theta,t)&=&\gamma_0(t)A(r)e^{iL\theta}\nonumber\\
&+&\sum_l\left[u_l(r,t)e^{il\theta}+v_l^\ast(r,t)e^{-il\theta}\right]e^{iL\theta}.
\end{eqnarray}
Conversely, $u_l$ and $v_l$ are written by using $\psi(r,\theta,t)$ as
\begin{subequations}
\begin{eqnarray}
 u_l(r,t)&=&\int_0^{2\pi}\frac{d\theta}{2\pi}\psi(r,\theta,t)e^{-i(L+l)\theta},\\
 v_l(r,t)&=&\int_0^{2\pi}\frac{d\theta}{2\pi}\psi^\ast(r,\theta,t)e^{i(L-l)\theta}.
\end{eqnarray}
\end{subequations}
We expand $\vect{w}_l(r,t)$ in eigenmodes of Bogoliubov equation as
\begin{eqnarray}
 \vect{w}_l(r,t)&=&\sum_{\lambda={\sf u,v}}\sum_{n=0,1,2,\cdots}\gamma_{ln}^{(\lambda)}(t)\vect{w}_{ln}^{(\lambda)}(r).
\end{eqnarray}
By using the orthonormal conditions \eqref{eq:all_orthonorm}, the coefficient is given by
$\gamma_{ln}^{\sf (u,v)}(t)=\pm \integ{\vect{w}_{ln}^{\sf (u,v)}(r)\hat{\sigma}\vect{w}_l(r,t)}$,
where the upper (lower) sign is for a superscript {\sf u} ({\sf v}).

In the extent of the linear approximation, 
$\gamma_{ln}^{\sf (u,v)}$ for a RE mode oscillates with the eigenfrequency $\omega_{ln}^{\sf (u,v)}$
as $\gamma_{ln}^{\sf (u,v)}\propto\exp(-i\omega_{ln}^{\sf (u,v)} t)$.
Then the amplitude is constant.
For a conjugate pair of CE modes $\frac{1}{\sqrt{2}}(\vect{w}_{ln'}^{\sf (u)}\pm i\vect{w}_{lm'}^{\sf (v)})$
with eigenfrequencies $\omega_c$ and $\omega_c^\ast$,
their time dependence are given by
$\frac{1}{\sqrt{2}}(\gamma_{ln'}^{\sf (u)}\pm i\gamma_{lm'}^{\sf (v)})\propto\exp[-i({\rm Re}\,\omega_c) t\mp({\rm Im}\,\omega_c) t]$, 
one of which is growing and the another is diminishing.

\subsection{Time development of vortex structures}
The solution of Eq.~\eqref{eq:GP-A} remains stationary without any perturbations.
We distort the trap potential in the $l_{\rm v}$-fold symmetrical form as
$V_{\rm tr}(r,\theta)=\frac{1}{4}r^2\left[1+\delta \cos(l_{\rm v} \ \!\theta)\right]$ during a short period.
By choosing the symmetry of the distortion,
the eigenmodes of $l=l_{\rm v}$ are selectively excited.

The distortion is small enough not to cause the distinguishable splitting of the multiply charged vortex.
When all of the excited modes have real eigenfrequencies,
the configuration of cores does not change in time.
In our simulation, no obvious splitting occurs at $\eta=500$ where no CE mode exists.

\begin{figure}
\includegraphics[width=8cm]{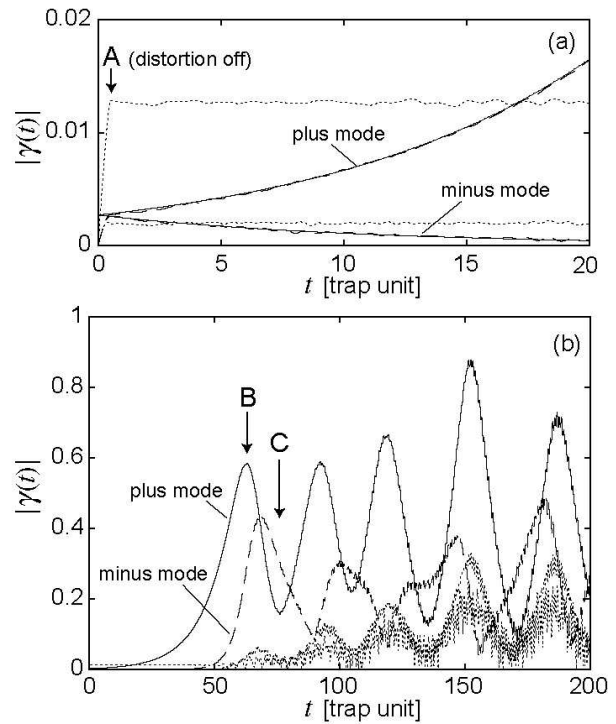}
\caption{\label{fig:excite3}Time development of elementary excitations.
(a) Initial changes.
The broken lines correspond to the amplitude of the CE modes (plus and minux modes),
which coincide with the solid lines corresponding to the functions $\exp(\pm{\rm Im}\,\omega_ct)$.
The amplitude of other modes plotted with the dotted lines are constant.
(b) Further time development.
%The two CE modes plotted with the solid and broken lines interact with each other and oscillate.
The plus and minus modes (the solid and broken lines, respectively) interact with each other and oscillate.
The amplitude of other modes (dotted lines) are also affected by the oscillation and increase more and more.}
\end{figure}

We have investigated how CE modes affect the configuration of vortices.
We have simulated at $\eta=730$
where exists a pair of CE modes $\frac{1}{\sqrt{2}}(\vect{w}_{33}^{\sf (u)}\pm i\vect{w}_{30}^{\sf (v)})$,
and distorted with $l_{\rm v}=3$.
Figure~\ref{fig:excite3} shows the time dependence of $\gamma_{3n}$,
which is the typical behavior of excitations including CE modes.
During all of the amplitudes of eigenmodes are small,
their time-dependence agrees well with the linear approximation as shown in Fig.~\ref{fig:excite3}(a):
the amplitude of one of the CE modes (plus mode) increases as $\exp({\rm Im}\omega_ct)$,
that of another CE mode (minus mode) decreases as $\exp(-{\rm Im}\omega_ct)$,
and those of RE modes are constant.
%the amplitudes of CE modes (broken lines) change as $\exp(\pm{\rm Im}\,\omega_c t)$ (solid lines),
%while others (dotted lines) are constant.
Further time development is shown in Fig.~\ref{fig:excite3}(b).
When the plus mode increases beyond the linear approximation, 
it stops growing and the minus mode starts growing instead.
They interact with each other and begin to oscillate.
Other excitations are also enlarged more and more throuth a non-linear process.
Here, RE modes of $l=l_{\rm v}$ are selectively enlarged.
(In the case of $l_{\rm v}=2$, we found the RE modes of $l=4$ are also excited through this non-linear process.)

\begin{figure}
\begin{minipage}{2.8cm}
(a)\\
\vspace{1mm}
\includegraphics[width=2.8cm]{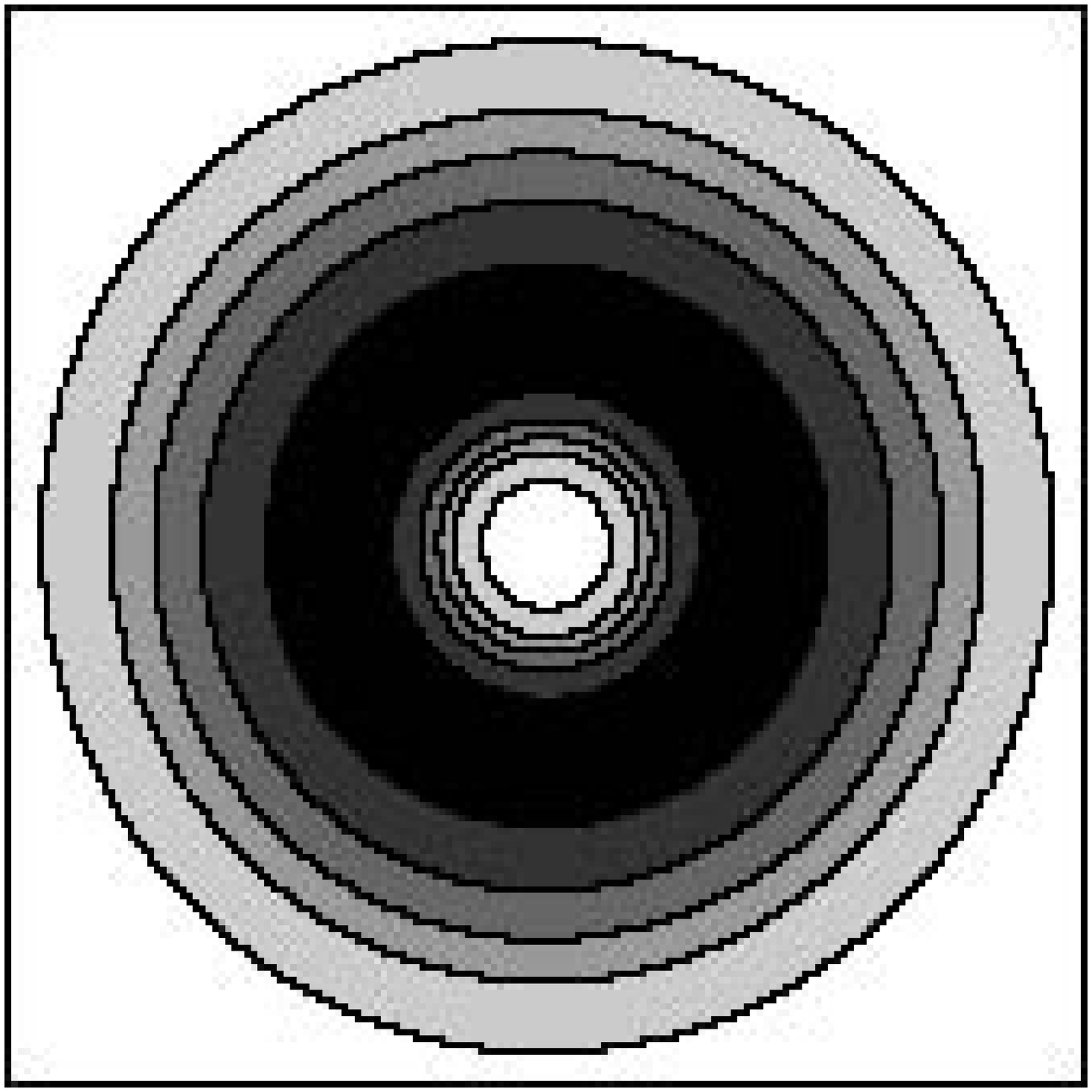}
\end{minipage}
\begin{minipage}{2.8cm}
(b)\\
\vspace{1mm}
\includegraphics[width=2.8cm]{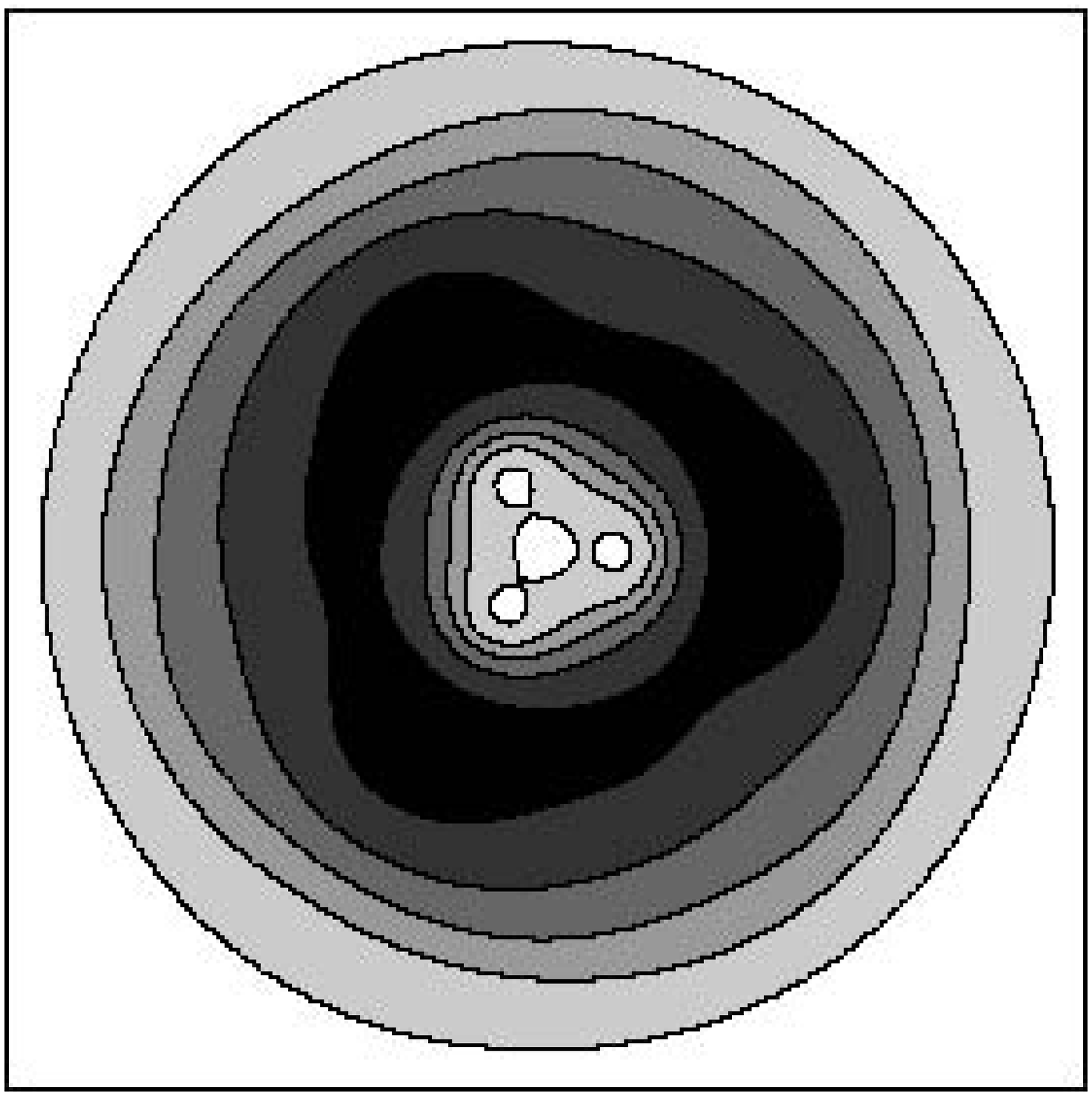}
\end{minipage}
\begin{minipage}{2.8cm}
(c)\\
\vspace{1mm}
\includegraphics[width=2.8cm]{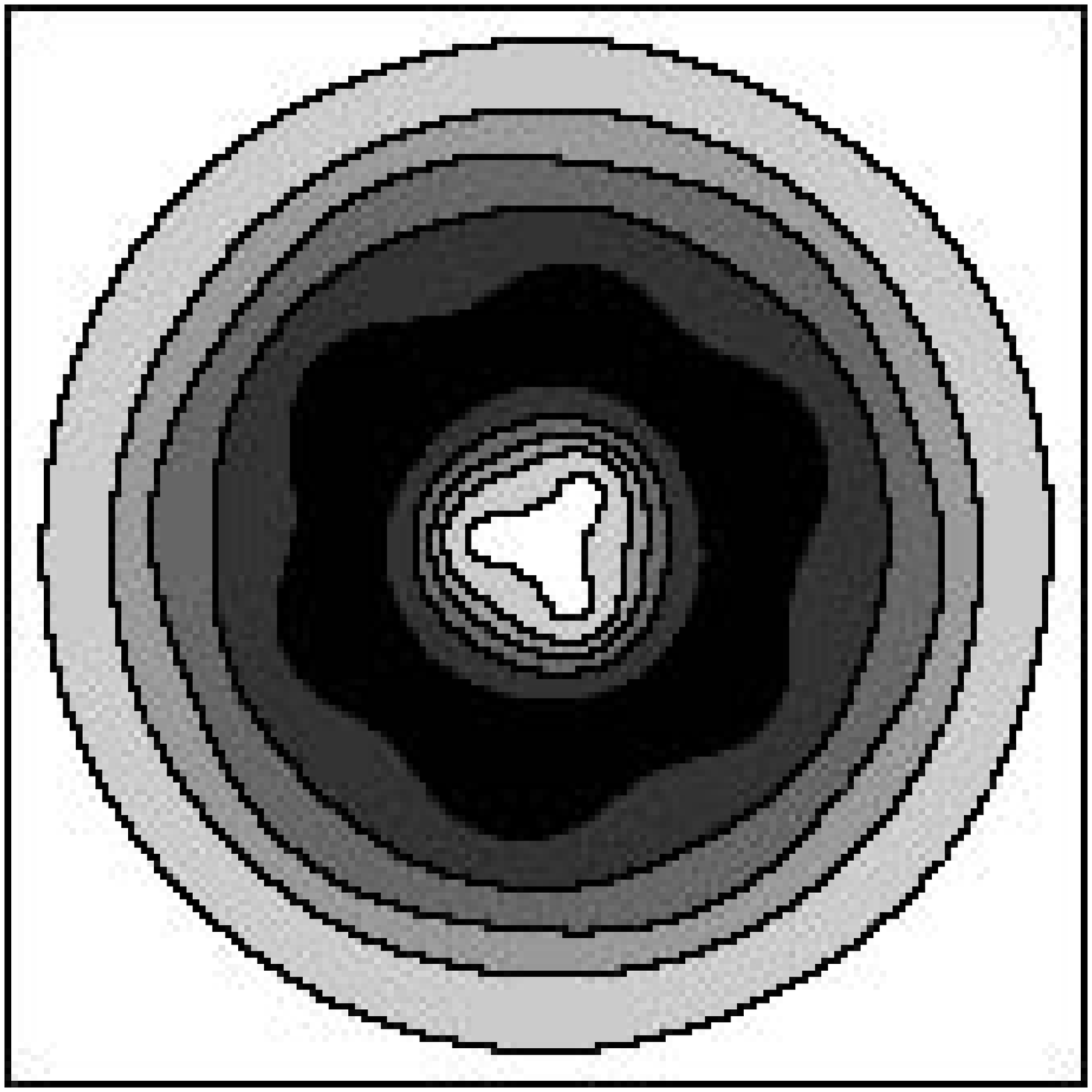}
\end{minipage}
\caption{\label{fig:dens3}Contour plots of density profiles at the points {\sf A, B} and {\sf C} in Fig.~\ref{fig:excite3}.
(a) A multiply charged vortex just after the distortion. (b), (c) Vortices arranged in three-fold symmetry.
Vortex cores move outward (b) and into the center (c) in the oscillation.}
\end{figure}

Figures~\ref{fig:dens3}(a), (b) and (c) are the images of condensates at the points {\sf A, B} and {\sf C} in Fig.~\ref{fig:excite3}, respectively.
The initial distortion causes little change of the distribution (Fig.~\ref{fig:dens3}(a)).
As the complex mode grows, the vortex split into four singly quantized vortices arranged in three fold symmetry (Fig.~\ref{fig:dens3}(b)).
The interaction of the CE modes results in the oscillation of distances among the cores.
The divided cores in Fig.~\ref{fig:dens3}(b) move toward the center again as shown in Fig.~\ref{fig:dens3}(c) and oscillate between these two states.
In the oscillation, singly charged vortices never return to the quadruply charged vortex.
Their average distance becomes larger and larger.
Simultaneously, the center core begins to vibrate and condensate becomes distorted more and more.
The energy of the multiply charged vortex is gradually transfered to the excitation energies of RE modes.

\begin{figure}
\begin{minipage}{2.8cm}
(a)\\
\vspace{1mm}
\includegraphics[width=2.8cm]{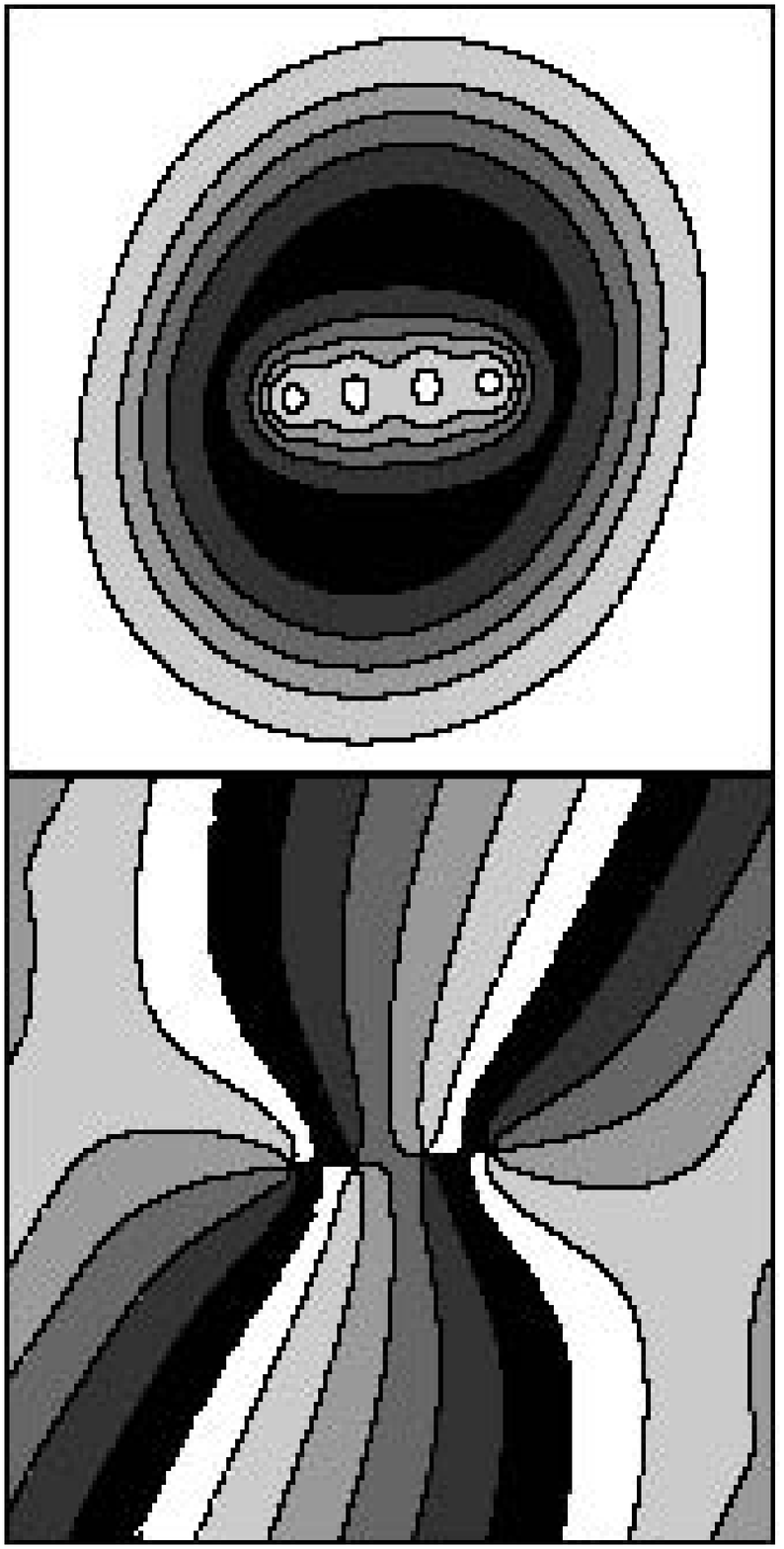}
\end{minipage}
\begin{minipage}{2.8cm}
(b)\\
\vspace{1mm}
\includegraphics[width=2.8cm]{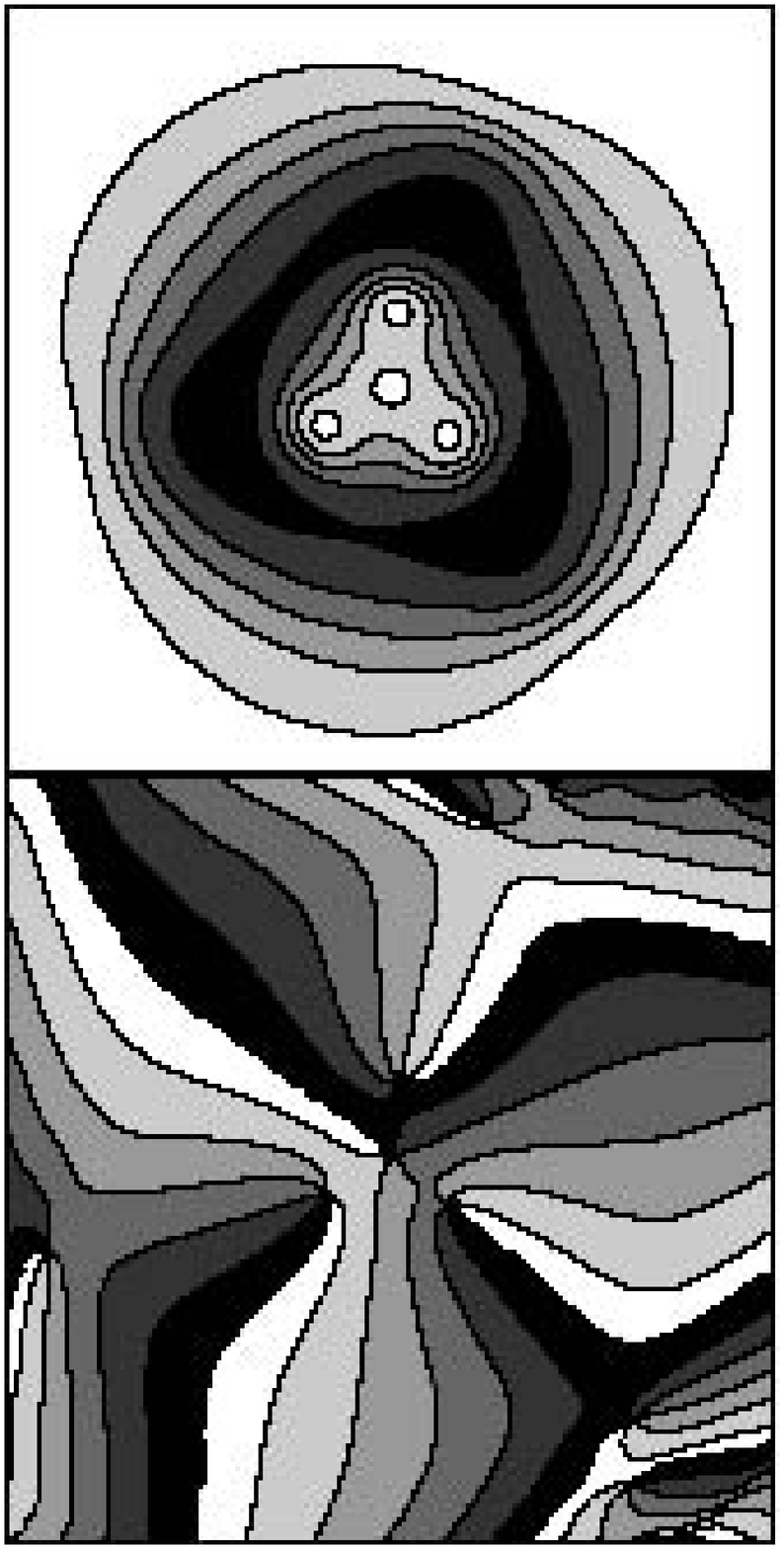}
\end{minipage}
\begin{minipage}{2.8cm}
(c)\\
\vspace{1mm}
\includegraphics[width=2.8cm]{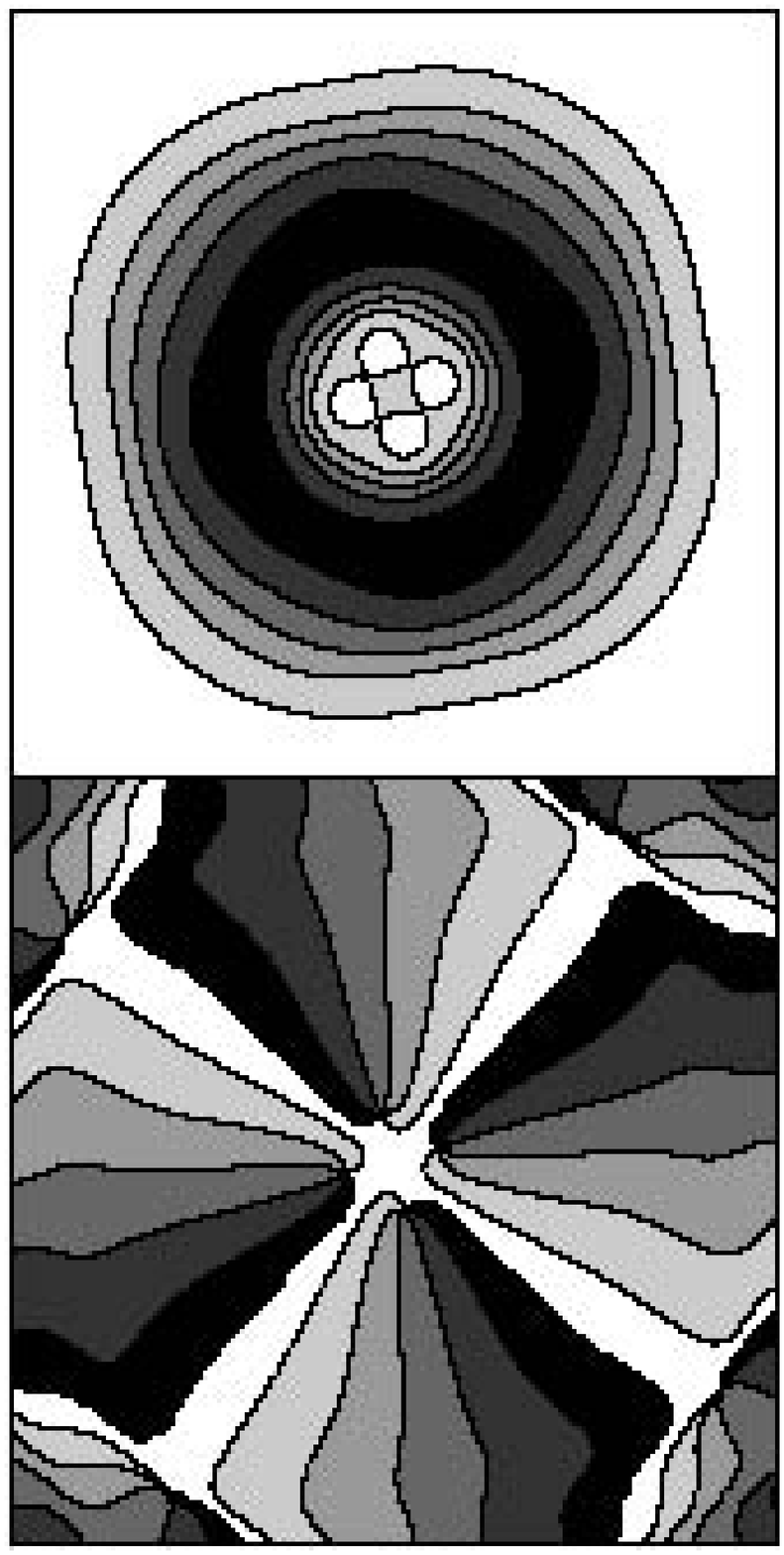}
\end{minipage}
\caption{\label{fig:d2}Images of splitting patterns at $\eta=340$ after trap distortion
 with (a) $l_{\rm v}=2$, (b) $l_{\rm v}=3$, and (c) $l_{\rm v}=4$.
The uppers are density profiles and the bottoms are corresponding phase profiles.
Three CE modes are selectively excited by choosing $l_{\rm v}$.}
\end{figure}

We have also investigated the coexistence region of several CE modes.
We have simulated at $\eta=340$ where CE modes of $l=2,3$ and $4$ are exist.
Figures~\ref{fig:d2}(a), (b) and (c) are the images of splitting after distorted with $l_{\rm v}=2, 3$ and $4$, respectively.
Vortices in each image are arranged in $l_{\rm v}$-fold symmetry. 
This fact means that the splitting patterns can be controlled by the initial distortion.

As we have mentioned before,
the symmetry of the configuration of vortices agrees with $l_{\rm v}$, i.e. the angular momentum of excited states.
At the center of the trap, however, 
a doubly quantized vortex, which appears in $l=2$ splitting, seems unstable
and soon decays into two singly charged vortices (Fig.~\ref{fig:d2}(a)).

\section{conclusion}
We have calculated the collective excitation spectra of a BEC with a quadruply charged vortex,
and discussed the origin of CE modes.
A CE mode is decomposed into two excitations, whose total energy and angular momentum are equal to zero.
This fact allows the CE mode to grow without any dissipative processes.
To satisfy the energy conservation, one of the elements must be a NE mode.
Its $\eta$-dependence affects the appearance of the CE mode.
We have found all of the CE modes existing in the region $0\le \eta \le 4000$ (Fig.~\ref{fig:im}).

The possible patterns of vortex splitting are classified by the angular momenta of fluctuations.
These structures have been confirmed by the numerical simulations of splitting.
Moreover, the selection of the patterns has found to be possible by the manipulation of the trap symmetry.
The intrinsic splitting are caused by a CE mode,
and if one choose the $\eta$ (i.e. the number of trapped atoms) so that a CE mode exits,
the splitting corresponding to the CE mode occurs.
The CE mode is efficiently excited by controlling the symmetry of the trap,
even in case when several CE modes coexist.

In this paper, we have assumed a pancake-shaped BEC and taken a 2-dimensional approach.
In a cigarette-shaped BEC as in the experiments, 3D simulation shows that the splitting cannot be recognized by time-of-flight
because of the excitation along a vortex line~\cite{Mottonen2003}.
%We hope the observation of the various structures as shown in this paper somehow in experiments.

\begin{acknowledgments}
This work is supported by the Grant-in-Aid for the 21st Century COE ``Center for Diversity and Universality in Physics''
from the Ministry of Education, Culture, Sports, Science and Technology (MEXT) of Japan.
%This work is supported by a Grant-in-Aid for the 21st Century COE ``Center for Diversity and Universality in Physics''.
We are grateful to K.~Machida, M.~Nakahara, T.~Mizushima, and T.~Ishoshima for useful discussions.
\end{acknowledgments}

%\newpage %Just because of unusual number of tables stacked at end
%\bibliography{sim}% Produces the bibliography via BibTeX.

\end{document}